\documentclass[aps,showpacs,11pt,superscriptaddress,preprintnumbers,amsmath,amssymb,nofootinbib]{revtex4-1}
\usepackage{CJK,upgreek,fancyhdr}
\pdfoutput=1
\usepackage{mathrsfs}
\usepackage{epsfig}
\usepackage{graphicx}
\usepackage{dcolumn}
\usepackage{bm}
\usepackage{amsmath}
\usepackage{slashed}       
\usepackage{amssymb}
\usepackage{amsfonts}
\usepackage{multirow}
\usepackage{makecell}
\usepackage{enumerate}
\usepackage[marginal]{footmisc}
\usepackage{float}
\usepackage{setspace}

\usepackage{graphics, setspace}

\usepackage{color,xcolor,fancybox,epsf,rotating,colordvi}
\usepackage{ulem}

\newcommand{\mathsym}[1]{{}}
\newcommand{\unicode}[1]{{}}

\newcommand{\PreserveBackslash}[1]{\let\temp=\\#1\let\\=\temp}
\newcolumntype{C}[1]{>{\PreserveBackslash\centering}p{#1}}
\newcolumntype{R}[1]{>{\PreserveBackslash\raggedleft}p{#1}}
\newcolumntype{L}[1]{>{\PreserveBackslash\raggedright}p{#1}}

\let\jnfont=\rm
\def\NPB#1,{{\jnfont Nucl.\ Phys.\ B }{\bf #1},}
\def\PLB#1,{{\jnfont Phys.\ Lett.\ B }{\bf #1},}
\def\EPJC#1,{{\jnfont Eur.\ Phys.\ Jour.\ C }{\bf #1},}
\def\PRD#1,{{\jnfont Phys.\ Rev.\ D }{\bf #1},}
\def\PRL#1,{{\jnfont Phys.\ Rev.\ Lett.\ }{\bf #1},}
\def\MPLA#1,{{\jnfont Mod.\ Phys.\ Lett.\ A }{\bf #1},}
\def\JPG#1,{{\jnfont J.\ Phys.\ G}{\bf #1},}
\def\CTP#1,{{\jnfont Commun.\ Theor.\ Phys.\ }{\bf #1},}
\def\ZPC#1,{{\jnfont Z.\ Phys.\ C }{\bf #1},}
\def\JHEP#1,{{\jnfont JHEP \ }{\bf #1},}
\def\lsim{\raise0.3ex\hbox{$<$\kern-0.75em\raise-1.1ex\hbox{$\sim$}}}
\def\gsim{\raise0.3ex\hbox{$>$\kern-0.75em\raise-1.1ex\hbox{$\sim$}}}

\newcommand{\GeV}{~\rm GeV}

\newcommand{\fbm}{{~\rm fb}^{-1}}
\newcommand{\fb}{~\rm fb}

\begin{document}
\begin{CJK*}{GBK}{song}
\begin{spacing}{1.0}

\title{A Light Scalar in the Minimal Dilaton Model in Light of LHC Constraints}
\author{Lijia Liu}
\affiliation{Center for Theoretical Physics, School of Physics and Technology, Wuhan University, Wuhan 430072, China}
\author{Haoxue Qiao}
\affiliation{Center for Theoretical Physics, School of Physics and Technology, Wuhan University, Wuhan 430072, China}
\author{Kun Wang}
\affiliation{Center for Theoretical Physics, School of Physics and Technology, Wuhan University, Wuhan 430072, China}
\author{Jingya Zhu}
\affiliation{Center for Theoretical Physics, School of Physics and Technology, Wuhan University, Wuhan 430072, China}
\affiliation{Enrico Fermi Institute, University of Chicago, Chicago, IL 60637, USA}

\date{2018-11-26}

\begin{abstract}
Whether an additional light scalar exists is an interesting topic beyond the Standard Model (SM), while nowadays we do not know exactly physics beyond the SM in the low mass region,
e.g., the Atlas and CMS collaborations get inconsistent results at around 95 GeV in searching for light resonances in diphoton channel.
Considering these, we study a light scalar in the Minimal Dilaton Model (MDM).
Under the theoretical and latest experimental constraints, we sort the surviving samples into two scenarios according to the diphoton rate of the light scalar:
the large-diphoton scenario (with $\sigma_{\gamma\gamma}/SM\gtrsim0.2$) and the small-diphoton scenario (with $\sigma_{\gamma\gamma}/SM\lesssim0.2$), which are favored by CMS and Atlas results respectively.
We compare the two scenarios, check the characteristics in model parameters, scalar couplings, production and decay, and consider further distinguishing them at colliders.
Finally, we get the following conclusions for the two scenarios:
(i) The formal usually has small Higgs-dilaton mixing angle ($|\sin\theta_S|\lesssim0.2$) and small dilaton vacuum expectation value (VEV) $f$ ($0.5\lesssim\eta\equiv v/f\lesssim1$), and the later usually has large mixing ($|\sin\theta_S|\gtrsim0.4$) or large VEV ($\eta\equiv v/f\lesssim0.3$).
(ii) The former usually predicts small $s\gamma\gamma$ coupling ($|C_{s\gamma\gamma}/SM|\lesssim0.3$) and large $sgg$ coupling ($0.6\lesssim|C_{sgg}/SM|\lesssim1.2$),
while the later usually predicts small $sgg$ coupling ($|C_{sgg}/SM|\lesssim0.5$).
(iii) The former can interpret the small diphoton excess by CMS at its central value, when $m_s\simeq95\GeV$, $\eta\simeq0.6$ and $|\sin\theta_S|\simeq0$.
(iv) The former usually predicts a negative correlation between Higgs couplings $|C_{h\gamma\gamma}/SM|$ and $|C_{hgg}/SM|$, while the later usually predicts the two couplings both smaller than 1, or $|C_{h\gamma\gamma}/SM|\lesssim0.9 \lesssim|C_{hgg}/SM|$.
\end{abstract}

\maketitle

\section{Introduction}
A 125 GeV Higgs was discovered at the LHC \cite{1207-a-com,1207-c-com}, with right spin and CP property, and 
production rates are consistent with the Standard Model (SM) globally according to both Atlas and CMS collaborations \cite{h-a-web, h-c-web, ATLAS-com2018, CMS-com2018, hfit-2018}.
While behind the deviations on Higgs production rates there still be chances for new physics, which can be different symmetry-breaking mechanisms, new particles in Higgs-coupling loops, or Higgs mixing with additional scalars.
And after Higgs discovered, whether an additional scalar exists is a natural question concerned most by both experimentalists and theorists.

However, we do not know exactly physics beyond the SM even in the low mass region.
Before the LHC, the largest center-mass energy at the LEP is 209 GeV, and it excluded a SM-like Higgs below 114.4 GeV finally before closed \cite{LEP2003}.
In fact, the data at LEP is so small that a light scalar can still be possible, with production rates below the SM prediction.
For example, recently the CMS collaboration presented their searches for low-mass new resonances decaying to two photons, both for the 8 TeV and 13 TeV data sets a small excess around 95 GeV was hinted at, with approximately 2.8 $\sigma$ local (1.3 $\sigma$ global) significance for a hypothesis mass of 95.3 GeV in the combined analysis \cite{CMS95}.
The signal around 95 GeV at the 13 TeV LHC is about $\sigma^{13TeV}_{\gamma\gamma}\approx80\pm20\fb$, or $\sigma^{13TeV}_{\gamma\gamma}/SM\approx0.64\pm0.16$.
Such a result was interpreted or discussed in several papers \cite{interpret95}.
In the same mass region and the same channel, the Atlas collaboration released their new search result with about $80\fbm$ data at 13 TeV, but no excess observed, with an excluded limit of $\sigma^{13TeV}_{\gamma\gamma}\lesssim60\fb$ at $95\%$ confidence level (CL) \cite{lightH-ATLAS2018}.
Compared with the CMS result, the signal around 95 GeV at the 13 TeV LHC by Atlas is about $\sigma^{13TeV}_{\gamma\gamma}\approx18\pm18\fb$, or $\sigma^{13TeV}_{\gamma\gamma}/SM\approx0.14\pm0.14$.
Considering the difference between these two collaborations, further checking that at the LHC or future colliders is necessary.
The difference between these two collaborations, together with the other small excess at 98 GeV \cite{LEP2003}, 28 GeV \cite{CMS28} and 115 GeV \cite{CMS115}, all reflect our unsureness of physics beyond the SM in the low mass region.
Thus it is still interesting to consider a light scalar in new physics models, which have different diphoton rates in different parameter spaces interpreting the results of the two  collaborations respectively, and to distinguish the parameter spaces at the LHC and future colliders.

In this letter, we consider this idea in the Minimal Dilaton Model (MDM),
which extends the SM by a dilaton-like singlet scalar and vector-like fermions \cite{mdm1, mdm2, 1311-MDM, 1405-MDM, MDM-750}.
Just like the traditional dilaton \cite{dilaton-before}, the singlet scalar in this model arises from a strong interaction theory with approximate scale invariance at a certain high energy scale, whose breakdown of the invariance triggers the electroweak symmetry breaking.
The singlet as the pseudo Nambu-Goldstone particle of the broken invariance, can be naturally light compared with the high energy scale.
Unlike traditional dilaton theory, this model assumes that only Higgs and top quark sectors, instead of all SM particles, can interact with the dynamics sector,
and consequently the singlet does not couple directly to the fermions and $W$, $Z$ bosons in the SM.
Meanwhile, the additional vector-like fermions acting as the lightest particles in the dynamical sector, to which the singlet naturally couples in order to recover the scale invariance: $M \to M e^{-\phi/f}$.
As a result, these fermions can induce the interactions between the pure singlet and the photons/gluons, or $Z$/$W$ boson with loop effect.
Furthermore, its mixing with the SM Higgs field can also induce the interactions.
Thus a light scalar can exist in MDM, mixed by the SM Higgs and singlet fields, and can be further checked at the LHC or future electron-positron colliders.
Due to the limitation of space in this letter, we leave the later checking study in our future work.

This letter is organized as follows. We first introduce briefly the MDM in Section II.
In Section III, we give the formulas for production rates of MDM scalars at the LHC.
In Section IV, we discuss the constraints to the model, and show the calculation and results.
Finally, we draw our conclusions in Section V.

\section{the Minimal Dilaton Model}
As introduced in Sec.\,I, the MDM extends the SM by a dilaton-like singlet field $S$ and a vector-like top partner field $T$. The effective Lagrangian can be written by  \cite{mdm1,mdm2}
\begin{eqnarray}
\mathcal L &=
	&	\mathcal L_{\rm SM}
		-\frac{1}{2}\partial_\mu S\partial^\mu S
		-\tilde V(S,H)\nonumber\\
	&&   -\overline T\left(\slashed{D}+\frac{M}{f}S\right)T
		-\left[y'\overline T_R(q_{3L}\cdot H)+{\rm h.c.}\right],  \label{lagr}
\end{eqnarray}
with $q_{3L}$, $M$ and $\mathcal L_{\rm SM}$ as the third-generation quark doublet, the strong-dynamics scale, and the SM Lagrangian without the Higgs potential respectively.
While the new scalar potential $\tilde{V}(S,H)$ can be generally given by
\begin{eqnarray}
\tilde V(S,H)
	&=&	M_H^2|H|^2+\frac{M_S^2}{2}S^2+\frac{\kappa}{2}S^2|H|^2
        +\frac{\lambda_H}{4}|H|^4+\frac{\lambda_S}{24}S^4,
		\label{scalar}
\end{eqnarray}
with $M_H$, $M_S$, $\kappa$, $\lambda_H$, and $\lambda_S$ as free parameters.
To break the symmetries, $H$ and $S$ get vacuum expectation values (VEVs) $v=246\GeV$ and $f$ respectively.
Then the singlet dilaton field $S$ can mix with the CP-even Higgs component $H^0$, forming two mass eigenstates $h,s$, that is
\begin{eqnarray}
\left[ \begin{array}{c} h \\ s \end{array} \right]
=
\left[ \begin{array}{cc}
\cos\theta_S & \sin\theta_S \\
-\sin\theta_S & \cos\theta_S
\end{array}
\right]
\left[ \begin{array}{c} H^0-\frac{v}{\sqrt{2}}\\ S-f \end{array} \right]
\end{eqnarray}
In this work we fix $m_h= 125.09\GeV$ which is the combined mass value of Atlas and CMS collaborations \cite{h1503-mass}.
For convenience we express $\lambda_H$, $\lambda_S$ and $\kappa$ by input parameters $f$, $v$, $\theta_S$, $m_h$ and $m_s$, and define \cite{mdm1,mdm2}
\begin{eqnarray}
  \eta &\equiv& \frac{v}{f} N_T ,
\end{eqnarray}
where $N_T$ is the number of field $T$, and we set it to 1 in this work.
Then under the conditions $m_{t'}\gg m_t$ and $\tan\theta_L\ll m_{t'}/m_t$, the
normalized couplings of $h$ and $s$ are given by \cite{mdm1,mdm2}
\begin{eqnarray}
C_{hVV}/SM	&=&	C_{hff}/SM	= \cos\theta_S,\nonumber\\
C_{sVV}/SM	&=&	C_{sff}/SM	= -\sin\theta_S,
\label{hffhvv}
\end{eqnarray}
where $V$ denotes either $W^{\pm}$ or $Z$ boson, and $f$ the fermions except for top quark sector.

The new fermion fields $(T_L,T_R)$ have the same quantum numbers with the SM fields $(q_{3L},u_{3R})$, thus they mix to form two mass eigenstates $(t,t')$, that is
\begin{eqnarray}
\left[ \begin{array}{c} t_L  \\ t'_L \end{array} \right]
= V_L^{\dagger}
\left[ \begin{array}{c} q_{3L}  \\ T_L \end{array} \right], \qquad
\left[ \begin{array}{c} t_R  \\ t'_R \end{array} \right]
= V_R^{\dagger}
\left[ \begin{array}{c} u_{3R}  \\ T_R \end{array} \right],
\end{eqnarray}
where we chose the mixing matrixes as
\begin{eqnarray}
V_L =
\left[ \begin{array}{cc}
\cos\theta_L & \sin\theta_L \\
-\sin\theta_L & \cos\theta_L \end{array}\right] ,\qquad
V_R =
\left[ \begin{array}{cc}
\cos\theta_R & \sin\theta_R \\
-\sin\theta_R & \cos\theta_R \end{array}\right].
\end{eqnarray}
From Eq.(\ref{lagr}), the mixing mass matrix is
\begin{eqnarray}
M_t=
\left[\begin{array}{cc}\frac{v}{\sqrt{2}} y_t  &\frac{v}{\sqrt{2}} y' \\
0 & f y_s \end{array} \right],
\end{eqnarray}
which then can be diagonalized as
\begin{eqnarray}
V_L^{\dagger} M_t V_{R} =
\left[ \begin{array}{cc} m_t & 0 \\0 & m_{t'} \end{array} \right].
\end{eqnarray}
Choosing $m_t,m_{t'},\theta_L$ as input parameters, then other parameters can be expressed as
\begin{eqnarray}
&&\tan\theta_R = \frac{m_t}{m_{t'}}\tan\theta_L, \nonumber\\
&&y_t = \frac{\sqrt{2}}{v}(m_t \cos\theta_L \cos\theta_R +m_{t'} \sin\theta_L \sin\theta_R)
= \frac{\sqrt{2}}{v}\frac{m_t m_{t'}}{\sqrt{m_t^2 \sin^2\theta_L +m_{t'}\cos^2\theta_L}},\nonumber\\
&&y' = \frac{\sqrt{2}}{v}(-m_t \cos\theta_L \sin\theta_R +m_{t'} \sin\theta_L \cos\theta_R)
= \frac{\sqrt{2}}{v}\frac{(m_{t'}^2 -m_t^2) \cos\theta_L \sin\theta_L}{\sqrt{m_t^2 \sin^2\theta_L +m_{t'}\cos^2\theta_L}},\nonumber\\
&&y_s = \frac{1}{f} \sqrt{m_t^2 \sin\theta_L^2 +m_{t'}\cos^2\theta_L} ~.
\label{yukawa}
\end{eqnarray}
Since gluon/photon can only couple to a pair of the same mass eigenstates $t$ or $t'$ at tree level, in calculations of loop-induced coupling of scalars to $gg/\gamma\gamma$, we can normalize Yukawa couplings of scalars to top quark sector to their SM values
\begin{eqnarray}
C_{ht\bar{t}}/SM &=& \cos^2\theta_L\cos\theta_S +\eta\sin^2\theta_L\sin\theta_S, \nonumber\\
C_{ht'\bar{t'}}/SM &=& \sin^2\theta_L\cos\theta_S +\eta\cos^2\theta_L\sin\theta_S, \nonumber\\
C_{st\bar{t}}/SM &=& -\cos^2\theta_L\sin\theta_S +\eta\sin^2\theta_L\cos\theta_S,
\nonumber\\
C_{st'\bar{t'}}/SM &=& -\sin^2\theta_L\sin\theta_S +\eta\cos^2\theta_L\cos\theta_S,
\label{htt}
\end{eqnarray}
while we should know there are new couplings of $h$, $s$ and $Z$ to a pair of different mass eigenstates $t$ and $t'$.
And with Eqs. (\ref{hffhvv}) and (\ref{htt}), we can get the normalized loop-induced couplings
\begin{eqnarray}
C_{hgg}/SM	&=&	\cos\theta_S +\eta\sin\theta_S,\nonumber\\
C_{h\gamma\gamma}/SM &=&\cos\theta_S -0.27\times\eta\sin\theta_S,\nonumber\\
C_{sgg}/SM	&=&	[-A_b \sin\theta_S
+A_t \times (-\cos^2\theta_L\sin\theta_S +\eta\sin^2\theta_L\cos\theta_S)
\nonumber\\
&&+A_{t'} \times (-\sin^2\theta_L\sin\theta_S +\eta\cos^2\theta_L\cos\theta_S)] /(A_t+A_b),
 \nonumber\\
C_{s\gamma\gamma}/SM &=&
[-(A_W+\frac{1}{3}A_b) \times \sin\theta_S
+\frac{4}{3}A_t \times (-\cos^2\theta_L\sin\theta_S +\eta\sin^2\theta_L\cos\theta_S)
\nonumber\\
&&+\frac{4}{3}A_{t'} \times (-\sin^2\theta_L\sin\theta_S +\eta\cos^2\theta_L\cos\theta_S) ]/[A_W+\frac{4}{3}A_t+\frac{1}{3}A_b].
\label{hgg}
\end{eqnarray}
where $A_i$ is the loop function presented in Refs.\cite{hsm-rev} with particle $i$ running in the loop.
When $m_s=95\GeV$, the loop-induced coupling $sgg, s\gamma\gamma$ can be approximated by
\begin{eqnarray}
C_{sgg}/SM &\simeq& -\sin\theta_S +\eta\cos\theta_S,\nonumber\\
C_{s\gamma\gamma}/SM &\simeq& -\sin\theta_S -0.31\eta\cos\theta_S
\label{sgg-coup}
\end{eqnarray}

\section{Production rates of MDM scalars at colliders}
In the MDM, we assumed $h$ as the 125 GeV Higgs.
Since current Higgs data of production rates are very like these of SM Higgs globally, the mixing angle $\theta_S$ between Higgs and dilaton can be very small \cite{1311-MDM}.
In this work, we consider the dilaton-like scalar $s$ being lighter, e.g., $65\sim122\GeV$, which can be constrained by low mass resonance searches at the LHC \footnote{for scalar lighter than 65 GeV, we checked that $|\sin\theta_S|$ are constrained to be very small by the inclusive Higgs search results at the LEP, and $\eta$ can be very large because there are no diphoton data at the LHC to constrain it.};
or 95 GeV, to interpret the suspected low-mass resonance by CMS collaboration.
Furthermore, we suggest to further check the light scalar at the LHC and future electron-positron colliders.
As we can foresee, the lighter scalar with mass about 95 GeV and small $\theta_S$ mainly decay into $gg$, $\gamma \gamma$, $f \bar{f}$ (such as $b \bar{b}$, $c \bar{c}$, and $\tau^+\tau^-$) \cite{LHCHXSWG}.
In this section, we list the formulae we used for the production and decay of the two scalars.

First, we list the decay and production information of a SM Higgs at 125 and 95 GeV respectively, which are taken from Ref.\cite{LHCHXSWG}.
In Tab.\ref{tab1}, we list the branching ratios and total width.
In Tab.\ref{tab2}, we list the cross sections at 13 TeV LHC, which are calculated at NNLO level.
\begin{table}[!tbh]
\caption{The decay branching ratios and the total width of a SM Higgs at 125 and 95 GeV respectively \cite{LHCHXSWG}.}
\label{tab1}
\renewcommand\arraystretch{1.2}
\begin{tabular}{|c|c|c|c|c|c|c|c|c|c|}
\hline
$M_H$ (GeV) & $b\bar{b}$ & $c\bar{c}$ & $\tau^+\tau^-$ & $WW^*$ & $ZZ^*$ & $gg$ & $\gamma\gamma$ & $\Gamma^{SM}_{tot}$ (MeV) 
\\ \hline
125.0 & 0.591 & 0.0289 & 0.0635 & 0.208 & 0.0262 & 0.0782 & 0.00231 & 4.07 
\\\hline
95.0 & 0.810 & 0.0397 & 0.0824 & 0.00451 & $0.000651$ & 0.0608 & 0.00141 & 2.38 
\\\hline
\end{tabular}
\caption{The production cross sections at 13-TeV LHC of a SM Higgs at 125 and 95 GeV respectively \cite{LHCHXSWG}. }
\label{tab2}
\renewcommand\arraystretch{1.2}
\begin{tabular}{|c|c|c|c|c|c|}
\hline
$M_H$ (GeV) & $\sigma^{SM}_{ggF}$ (pb) & $\sigma^{SM}_{VBF}$ (pb) & $\sigma^{SM}_{WH}$  (pb) &  $\sigma^{SM}_{ZH}$ (pb) &  $\sigma^{SM}_{Ht\bar{t}}$ (pb)
\\ \hline
125.0 & 43.92 & 3.748 & 1.380 & 0.8696 & 0.5085
\\ \hline
95.0 & 70.64 & 3.680 & 2.931 & 1.622 & 0.5349
\\\hline
\end{tabular}
\end{table}

With the decay information for SM Higgs, the total width and branching ratios of the scalars $\phi=h,s$ in MDM can be written as
\begin{eqnarray}
\Gamma_{tot}^{\phi} &=& \Gamma^{SM}_{tot} \times \sum_{xx} \left[ Br^{SM}_{\phi\to xx} \times |C_{\phi xx}/SM|^2, \label{tot} \right] \\
Br_{\phi\to xx} &=& Br^{SM}_{\phi\to xx} \times |C_{\phi xx}/SM|^2 \times \frac{\Gamma^{SM}_{tot}}{\Gamma_{tot}^{\phi}},
\end{eqnarray}
where $xx=b\bar{b}, c\bar{c},\tau^+\tau^-,WW^*,ZZ^*,gg,\gamma\gamma$, and $C_{\phi xx}/SM$ are the corresponding normalized Yukawa couplings at leading order defined in Eqs. (\ref{hffhvv}), (\ref{htt}) and (\ref{hgg}).

And with the production information for SM Higgs, the production rates of the scalars $\phi=h,s$ in MDM at 13 TeV LHC can be calculated as
\begin{eqnarray}
\sigma_{ggF} &=& \sigma^{SM}_{ggF}(m_\phi) \times |C_{\phi gg}/SM|^2, \\
\sigma_{VBF,V\phi} &=& \sigma^{SM}_{VBF,VH}(m_\phi) \times |C_{\phi VV}/SM|^2, \\
\sigma_{\phi t\bar{t}} &=& \sigma^{SM}_{H t\bar{t}}(m_\phi) \times |C_{\phi t\bar{t}}/SM|^2,
\end{eqnarray}
where $C_{\phi xx}/SM$ with $xx=gg,WW,ZZ,t\bar{t}$, are also the corresponding normalized Yukawa couplings at leading order, defined in Eqs.(\ref{hgg}), (\ref{hffhvv}), and (\ref{htt}).

From the formulas and information above, one can get the following important conclusions:
\begin{itemize}
\item When $|\tan\theta_S| \gg \eta/4$, the dominated decay branching ratio of the light scalar of 95 GeV is $s\to b\bar{b}$, thus its total width and main decay branching ratios are
    \begin{eqnarray}
    \Gamma_{tot}^{s} \simeq 2.4 |\sin\theta_S|^2 {\rm ~MeV},  \quad 
    Br_{s\to b\bar{b}} \simeq 0.8,  \quad 
    Br_{s\to gg} \simeq 0.06,  \quad 
    Br_{s\to \gamma\gamma} \simeq 0.0014,
    \end{eqnarray}
    Where the branching ratio of diphoton can be a little larger (smaller) when the $\tan\theta_S$ is positive (negative).
\item When $|\tan\theta_S| \ll \eta/4$, the dominated decay branching ratio of the light scalar of 95 GeV is $s\to gg$, thus its total width and main decay branching ratios are
    \begin{eqnarray}
    \Gamma_{tot}^{s} \simeq 0.15 \eta^2 {\rm ~MeV},  \qquad 
    Br_{s\to gg} \simeq 1,  \qquad 
    Br_{s\to \gamma\gamma} \simeq 0.0022,
    \end{eqnarray}
    Where the branching ratio of diphoton can be a little larger (smaller) when the small $\tan\theta_S$ is positive (negative).
\item When $|\tan\theta_S|$ or $|\sin\theta_S|$ is small, the production rate of $s$ at the LHC can be proportional to $\eta^2$.
    Thus the golden probing channel for light dilaton at the LHC will be $gg\to s \to \gamma\gamma$, whose cross section can be approximated by
    \begin{eqnarray}
    \sigma_{\gamma \gamma} (m_s) \simeq \eta^2 \times \sigma^{SM}_{ggF}(m_s)  \times Br_{s\to\gamma\gamma},
     \label{diphoton}
    \end{eqnarray}
\item When $|\sin\theta_S|$ or $|\tan\theta_S|$ is not small, the vector bosons fusion (VBF) and vector boson scalar strahlung (Vs) production rates can be significant at tree level, and $s\to b\bar{b}$ with $s$ produced through VBF or Vs can be served as another channel to check it at the LHC, whose cross section can be approximated by
    \begin{eqnarray}
    \sigma_{VBF,b\bar{b}} (m_s) \simeq |\sin\theta_S|^2 \times \sigma^{SM}_{VBF}(m_s) \times Br_{s\to b\bar{b}} ~, \nonumber\\
    \sigma_{Vs,b\bar{b}} (m_s) \simeq |\sin\theta_S|^2 \times \sigma^{SM}_{VH}(m_s) \times Br_{s\to b\bar{b}} ~.
    \end{eqnarray}
    For this case, it can also be checked at future electron-positron colliders.
\item When $|\sin\theta_S|\approx0$, the loop effect of $t/t'$ in the effective coupling of $sZZ$ may be non-ignorable \cite{HZZ-loop, HZZ-Kniehl1990, H-COUP}.
    We leave this study in our future work.
\end{itemize}

\section{Numerical results and discussions}
In this section, we first scan over the parameter space of MDM under various experimental constraints.
Then for the surviving samples, we investigate the features of $h$ and $s$.
Before our scan, we clarify the following facts
\begin{itemize}
\item Firstly, since the characters of the dilaton in the MDM differ greatly from these of the SM Higgs boson, its mass may vary from several GeV to several hundred GeV without conflicting with LEP and LHC data in searching for Higgs boson.
    In fact, both Atlas and CMS released their results searching for low mass resonances in the region of 65-122 GeV at the LHC respectively \cite{CMS95, CMS95-2015, lightH-ATLAS2018, lightH-ATLAS2014}.
\item Secondly, since the diphoton rate of the light scalar is constrained by LHC data, $\eta\equiv v/f$ cannot be very large, thus we take $0<\eta\leq10$, and pay special attention to the case $\eta < 1$ in our study.
\item Thirdly, although in principle $\theta_S$ may vary from $-\pi/2$ to $\pi/2$, the Higgs data have required it to be around zero so that $h$ is mainly responsible for the electroweak symmetry breaking. In practice, requiring $|\tan\theta_S| \leq 2$ will suffice.
\item Finally, in MDM $t'$ may decay into $th,ts,tZ,bW$ at tree level. With $36\fbm$ data at 13-TeV LHC, the combined analyses of $t'\to tH,tZ,bW$ by Atlas excluded a vector-like $t'$ below 1.31 TeV at $95\%$ CL \cite{Atlas-VLQ}.
    While that of CMS exclude $t'$ with masses below 1140-1300 GeV \cite{CMS-VLQ}.
    The perturbativity may also require $t'$ not too heavy.
\end{itemize}
With the above considerations, we first scan the following parameter space:
\begin{eqnarray}
  && 0.01< \eta <10, ~~~~ |\tan\theta_S|<2, ~~~~ 0< |\sin\theta_L| <1, \nonumber\\
  && 65~{\rm GeV} <m_s< 122~{\rm GeV}, ~~~~ 1~{\rm TeV} <m_{t'}< 100~{\rm TeV}.
\end{eqnarray}
In our scan, we consider the following theoretical and experimental constraints:
\begin{itemize}
\item[(1)] Theoretical constraint of vacuum stability for the scalar potential, which corresponds to the requirement $4 \lambda_H \lambda_S - \lambda_{HS}^2 >0$ \cite{mdm1}.
\item[(2)] Theoretical constraints of perturbativity for scalar couplings $\lambda_S, \lambda_H, \kappa < 4 \pi$ and Yukawa couplings $y_t, y', y_s < 4 \pi$.
\item[(3)] Theoretical constraints from requiring that no Landau pole exists below 1 TeV.
    For the parameter running, we use the renormalization group equations (RGE) of the three scalar coupling parameters which are derived with \textsf{SARAH-4.12.3} \cite{SARAH},
    \begin{eqnarray}
      {\mathcal D} & \equiv & 16 \pi^2 \mu \frac{d}{d Q},  \nonumber\\
      {\mathcal D} \lambda_H &=& 6 \lambda_H^2 +2\kappa^2,  \nonumber\\
      {\mathcal D} \lambda_S &=& 3 \lambda_S^2 +12\kappa^2,  \nonumber\\
      {\mathcal D} \kappa &=& \kappa (3\lambda_H + \lambda_S) +4\kappa^2,
    \end{eqnarray}
\item[(4)] Experimental constraints from the electroweak precision data (EWPD). We calculate the Peskin-Takeuchi $S$ and $T$ parameters \cite{STU} with the formulae presented in \cite{mdm1}, and construct $\chi^2_{ST}$ by following experimental fit results with $m_{h,\rm{ref}}=125\GeV$ and $m_{t,\rm{ref}}=173\GeV$ \cite{PDG2018}:
    \begin{eqnarray}
        S=0.02\pm0.07, ~~ T=0.06\pm0.06, ~~ \rho_{ST}=0.92.
        \label{STdata}
    \end{eqnarray}
    In our calculation, we require that the samples satisfy $\chi^2_{ST} \leq 6.18$ \footnote{The data of $S$ and $T$ are from the global fit result to electroweak precision observables (EWPOs), mainly determinated by the EWPOs $Z$ boson mass $m_Z$ and width $\Gamma_Z$ (correlated with each other experimentally) respectively, so there is a strong correlation ($\rho_{ST}=0.92$) between the two parameters $S$ and $T$ \cite{PDG2018}.}. We do not consider the constraints from $V_{tb}$ and $R_b$ since they are weaker than these of the $S,T$ parameters \cite{mdm1}.
\item[(5)] Experimental constraints from the LEP, Tevatron and LHC search for Higgs-like resonances.
    We implement these constraints with the package $\textsf{HiggsBounds-5.2.0beta}$ \cite{HiggsBounds}.
    For the case we consider in this work ($65<m_s<122\GeV$, cross section and decay calculated at leading order), the main constraints to the light scalar come from diphoton results at the LHC \cite{CMS95, CMS95-2015, lightH-ATLAS2014, lightH-ATLAS2018}, and $Zb\bar{b}$ channel at the LEP \cite{LEP2003}
    \footnote{For the $Zjj$ channel at the LHC \cite{CMS115}, background is so large that the excluded limit is hundreds of times larger than the $ZH$ cross section; for the $Zjj$ and $Z\gamma\gamma$ channels at the LEP \cite{LEP-Zjj, LEP-Zgg}, $Zs$ production rate at tree level is anti-correlated with the $s\to gg,\gamma\gamma$ branching ratios; for the $\gamma s$ channels at the LEP \cite{LEP-gh}, the loop-induced $s\gamma\gamma$ and $sZ\gamma$ couplings are both very small. Thus all the existing results in these channels cannot give stronger constraints than the $Zb\bar{b}$ channel at the LEP \cite{LEP2003}.}.
\item[(6)] Experimental constraints from the 125 GeV Higgs data at LHC Run I and Run II.
    We first use the method in our former works \cite{1309-LightHiggs, 1311-MDM}, while with Higgs data updated with Figure 3 in \cite{15007-Atlas} and Figure 5 left in \cite{1412-CMS}.
    There are 20 experimental data sets in total, so we require $\chi^2_{20}\leq31.4$, which means each surviving sample fits 20 experimental data sets at $95\%$ CL \footnote{By this approach we only consider degrees of freedom in the experimental data, and we judge a model only by how well it can fit to the experimental data, without caring how many parameters in theoretical models. We think it is more objective by this approach since we do not know behind the data what the real theory is.}.
    Then we use $\textsf{HiggsSignal-2.2.0beta}$ \cite{HiggsSignal} which includes both Run I and Run II data.
    We require $\chi^2_{117}<143.2$, which means the P value with Higgs data $P_h>0.05$, or each surviving sample fits the total 117 experimental data sets at $95\%$ CL.
\end{itemize}
With samples satisfying all the constraints list above, we analyze their parameters, couplings and production rates of the scalars.

\begin{figure}[!htb]
  \centering
\includegraphics[width=10cm]{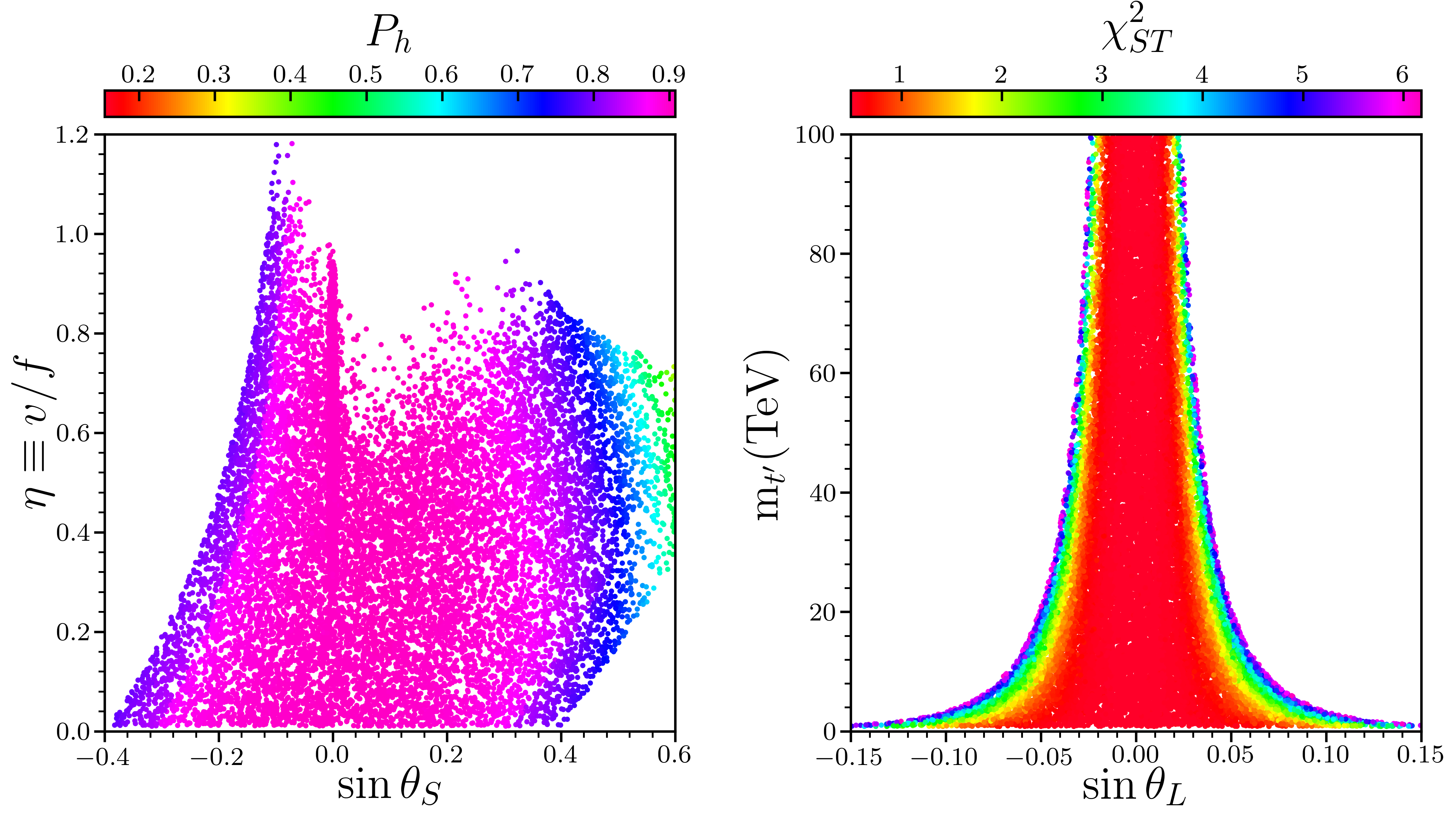}
\vspace*{-0.4cm}
\caption{Surviving samples on the $\eta \equiv v/f$ versus $\sin\theta_S$ (left), and $\sin\theta_L$ versus $m_{t'}$ (right) planes.
The colors indicate $P_h$ (left) and $\chi^2_{ST}$ (right) respectively, where $P_h$ is the P value calculated with the latest LHC Run-I and Run-II Higgs data in {\sf HiggsSignal-2.2.0}, and $\chi^2_{ST}$ is the $\chi^2$ in EWPD fit of parameters $S$ and $T$.
}
\label{fig1}
\end{figure}
In Fig.\ref{fig1}, we project the surviving samples on the planes of $\eta \equiv v/f$ versus $\sin\theta_S$ (left), and $m_{t'}$ versus $\sin\theta_L$ (right) respectively.
The colors indicate $P_h$ (left) and $\chi^2_{ST}$ (right) respectively, where $P_h$ is the P value calculated with the latest LHC Run-I and Run-II Higgs data in {\sf HiggsSignal-2.2.0}, and $\chi^2_{ST}$ is the $\chi^2$ in EWPD fit of parameters $S$ and $T$.
From this figure we can see that:
\begin{itemize}
\item Our strategy of Higgs fit is complementary with that of {\sf{HiggsSignal}}.
    Samples with $0.05<P_h\lesssim0.2$ are excluded by our strategy, while we checked that samples with $22\lesssim{\chi'}^2_h<31.4$ (or $0.05<P'_h\lesssim0.5$) in our strategy are excluded by {\sf{HiggsSignal}}.
\item According to {\sf HiggsSingnal}, the latest Higgs data, combined with constraints to the light scalar, can exclude samples with $\eta\gtrsim1$ or $|\sin\theta_S|\gtrsim0.5$, while these with $\eta\lesssim1$ and $|\sin\theta_S|\lesssim 0.3$ can fit the latest Higgs data at about $80\%\thicksim90\%$ level.
\item EWPD fit is very powerful in constraining the parameter $\sin\theta_L$, especially when the top partner is rather heavy.
    With $t'$ at 1 TeV $|\sin\theta_L|\gtrsim0.15$ is excluded, and with $t'$ at 20 TeV $|\sin\theta_L|\gtrsim0.05$ is excluded.
\end{itemize}

\begin{figure}[!htb]
  \centering
\includegraphics[width=15cm]{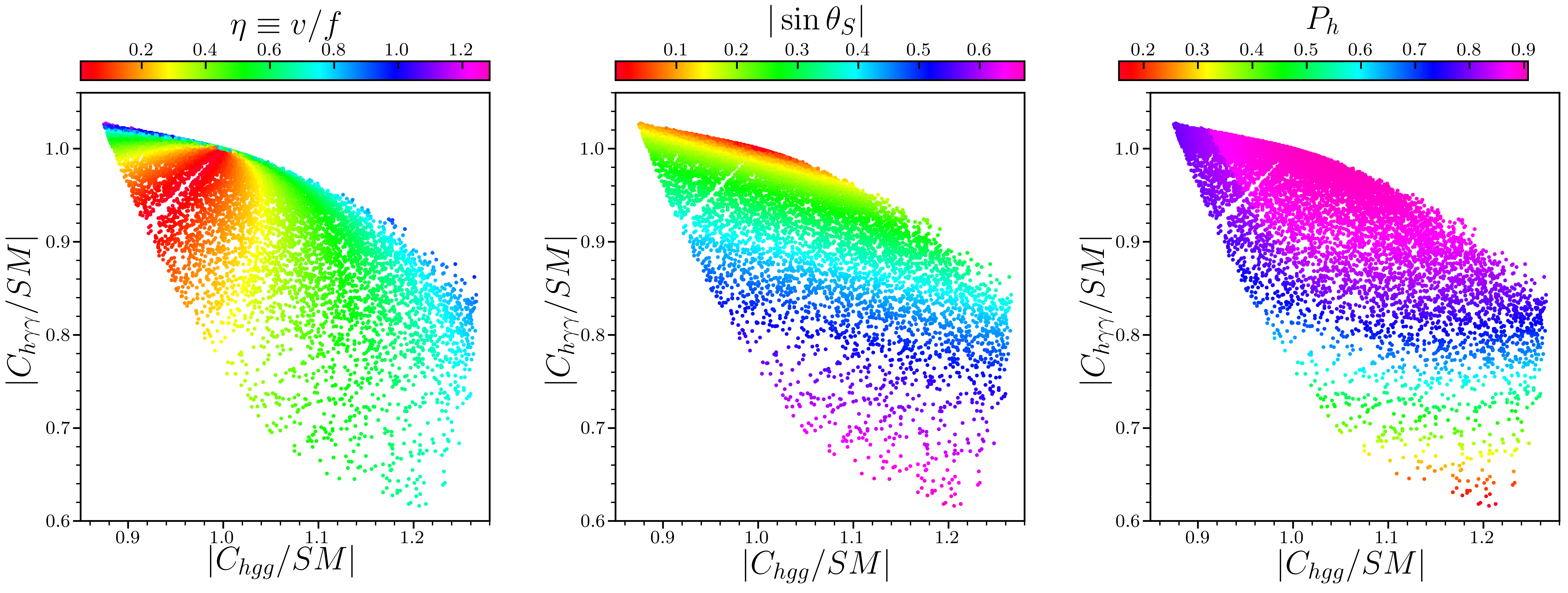}
\vspace*{-0.4cm}
\caption{Same samples as in Fig.\ref{fig1}, but on the planes of $|C_{h\gamma\gamma}/SM|$ versus $|C_{hgg}/SM|$, which are the normalized SM-like Higgs couplings to photons and gluons respectively.
The colors indicate $\eta$ (left), $|\sin\theta_S|$ (middle) and $P_h$ (right) respectively.}
\label{fig2}
\end{figure}
To interpret the Higgs fit result in Fig.\ref{fig1}, we project the surviving samples on the planes of $|C_{h\gamma\gamma}/SM|$ versus $|C_{hgg}/SM|$ in Fig.\ref{fig2}, with colors indicate $\eta$ (left), $|\sin\theta_S|$ (middle) and $P_h$ (right) respectively.
From this figure we can see that:
\begin{itemize}
  \item To fit Higgs data over $70\%$ level, the normalized coupling of the SM-like Higgs to gluons and photons should satisfy  $0.8\lesssim|C_{h\gamma\gamma}/SM|\lesssim1.05$ and $0.85\lesssim|C_{hgg}/SM|\lesssim1.25$.
  \item When $\eta\thicksim1$ there is negative correlation between $|C_{h\gamma\gamma}/SM|$ and $|C_{hgg}/SM|$, while when $\eta\lesssim0.3$ the two couplings are both smaller than 1.
  \item When $|\sin\theta_S|\lesssim0.2$ there is also negative correlation between $|C_{h\gamma\gamma}/SM|$ and $|C_{hgg}/SM|$, the relation is roughly  $C_{h\gamma\gamma}/SM \simeq 1.27-0.27\times C_{hgg}/SM$,
      while when $|\sin\theta_S|\gtrsim0.4$ $|C_{h\gamma\gamma}/SM|\lesssim0.9\lesssim|C_{hgg}/SM|$.
\end{itemize}

\begin{figure}[!htb]
  \centering
\includegraphics[width=10cm]{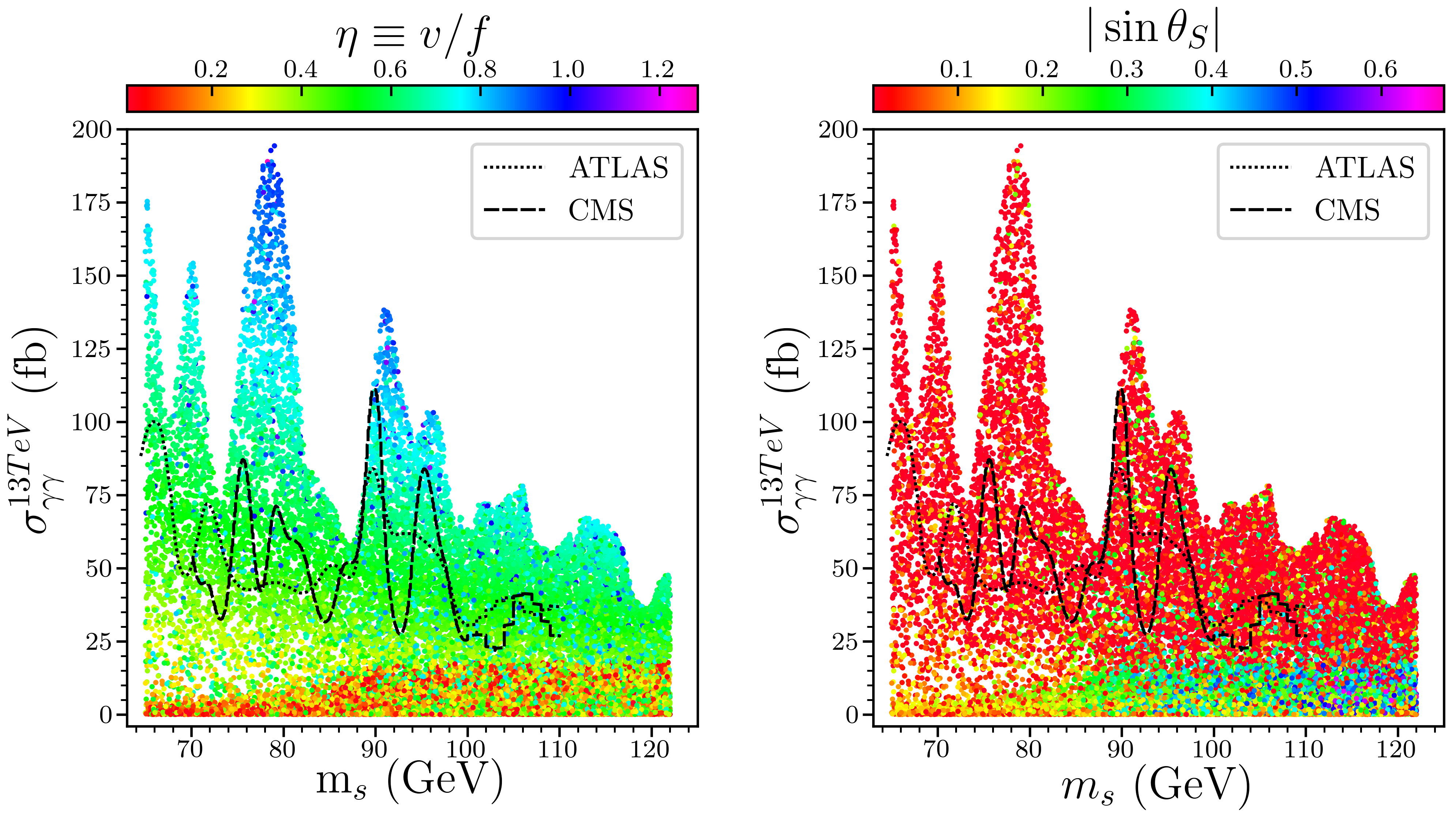}
\vspace*{-0.4cm}
\caption{Same samples as in Fig.\ref{fig1}, but on the planes of $\sigma^{13\rm TeV}_{\gamma\gamma}$ versus $m_{s}$, with colors indicate $\eta$ (left) and $|\sin\theta_S|$ (right) respectively.
The curves are the excluded limits in searching for low-mass resonance in diphoton channel at 13 TeV LHC, with dotted one by Atlas of $80\fbm$ \cite{lightH-ATLAS2018}, and dashed one by CMS of $35.9\fbm$ \cite{CMS95} respectively.}
\label{fig3}
\end{figure}
In Fig.\ref{fig3}, we project the surviving samples on the planes of $\sigma^{13\rm TeV}_{\gamma\gamma}$ versus $m_{s}$, with colors indicate $\eta$ (left) and $|\sin\theta_S|$ (right) respectively.
$\sigma^{13\rm TeV}_{\gamma\gamma}$ is the diphoton cross section of the light Higgs at 13-TeV LHC, and the dotted and dashed curves are the excluded limits by Atlas of $80\fbm$ data \cite{lightH-ATLAS2018}, and CMS of $35.9\fbm$ data \cite{CMS95} respectively.
We do not use this two excluded curves to constrain our samples, because the two results are inconsistent with each other at 95 GeV.
Instead we use the results of the two collaborations at Run I \cite{lightH-ATLAS2014, CMS95-2015} as the solid constraints.
According to the diphoton rate of the light scalar, we can roughly sort the MDM into two scenarios:
\begin{itemize}
  \item Large-diphoton scenario, which usually has small Higgs-dilaton mixing angle ($|\sin\theta_S|\lesssim0.2$) and small dilaton VEV $f$ ($0.5\lesssim\eta\equiv v/f\lesssim1$);
  \item Small-diphoton scenario, which usually has large Higgs-dilaton mixing angle ($0.4\lesssim|\sin\theta_S|\lesssim0.7$) or large dilaton VEV $f$ ($\eta\equiv v/f\lesssim0.3$).
\end{itemize}
We can see that the former scenario usually predict large diphoton rates ($\sigma^{13\rm TeV}_{\gamma\gamma}\gtrsim 20\fb$ or $\sigma^{13\rm TeV}_{\gamma\gamma}/SM\gtrsim 0.2$), while the later scenario usually predicts small diphoton rates ($\sigma^{13\rm TeV}_{\gamma\gamma}\lesssim 20\fb$ or $\sigma^{13\rm TeV}_{\gamma\gamma}/SM\lesssim 0.2$).
For the central value of CMS excess at $m_{\gamma\gamma}\simeq95\GeV$, $\sigma^{13\rm TeV}_{\gamma\gamma}/SM\simeq0.64$, we checked that samples of $m_s\simeq95\GeV$, $\eta\simeq0.6$ and $|\sin\theta_S|\simeq0$ can fit it well.

\begin{figure}[!htb]
  \centering
\includegraphics[width=10cm]{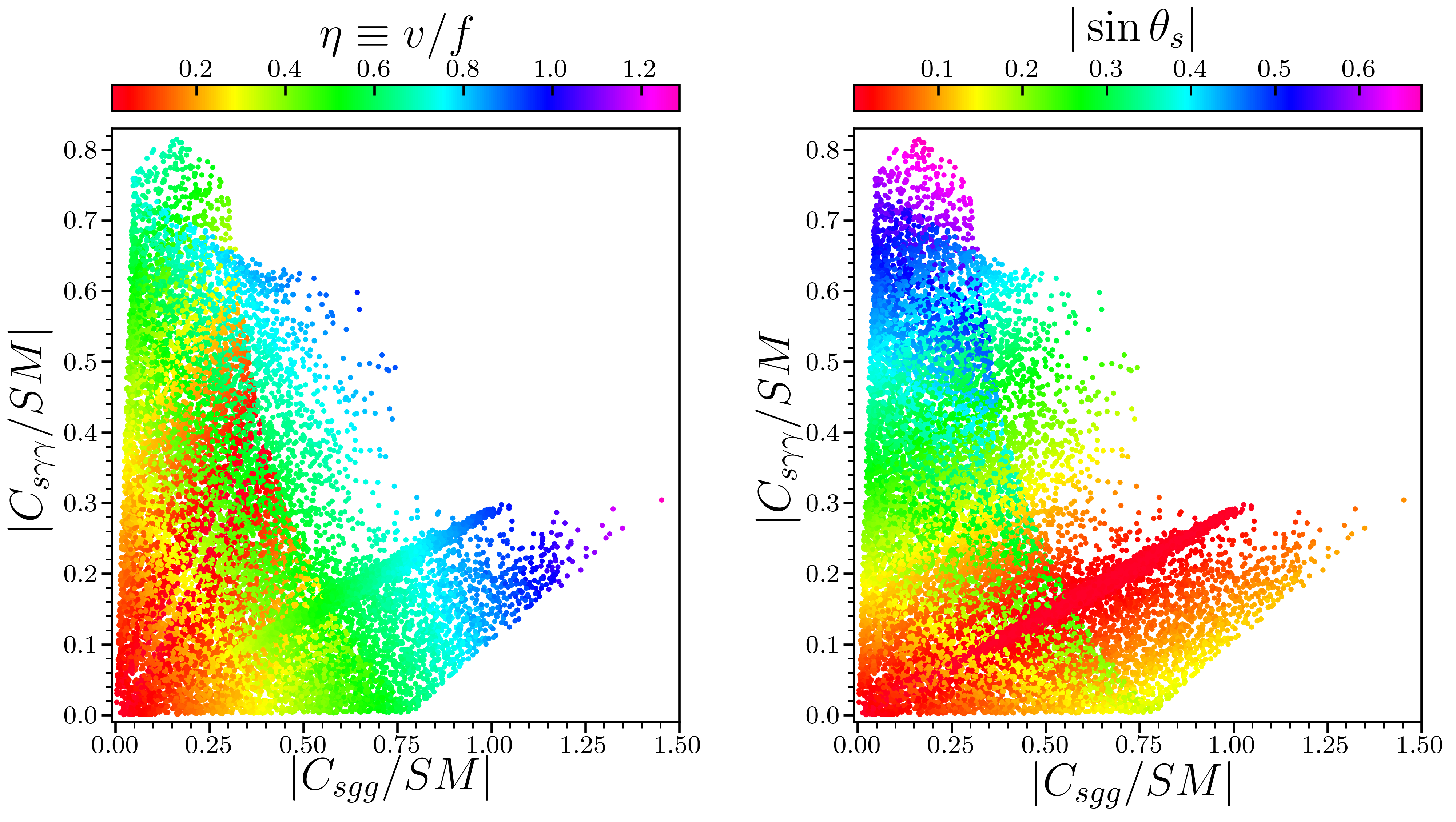}
\vspace*{-0.4cm}
\caption{Same samples as in Fig.\ref{fig1}, but on the planes of $|C_{s\gamma\gamma}/SM|$ versus $|C_{sgg}/SM|$, with colors indicate $\eta$ (left) and $|\sin\theta_S|$ (right) respectively.
The quantities $|C_{s\gamma\gamma}/SM|$ and $|C_{sgg}/SM|$ are the normalized light scalar couplings to photons and gluons respectively.
}
\label{fig4}
\end{figure}
To interpret the production rates of light Higgs in diphoton channel, in Fig.\ref{fig4} we project the surviving samples on the planes of $|C_{s\gamma\gamma}/SM|$ versus $|C_{sgg}/SM|$, with colors indicate $\eta$ (left) and $|\sin\theta_S|$ (right) respectively.
The quantities $|C_{s\gamma\gamma}/SM|$ and $|C_{sgg}/SM|$ are the normalized light Higgs coupling to gluons and photons respectively.
From this figure we can see that:
\begin{itemize}
  \item When $|\sin\theta_S|\thickapprox0$, we checked that the ratio of the two normalized  loop-induced couplings can be
      \begin{eqnarray}
      \frac{|C_{s\gamma\gamma}/SM|}{|C_{sgg}/SM|} \approx 0.3,
      \end{eqnarray}
      which can also be inferred from Eq.(\ref{sgg-coup}).
  \item Samples with small $|\sin\theta_S|$ and large $\eta$ have large $sgg$ couplings ($0.6\lesssim|C_{sgg}/SM|\lesssim1.2$) and small $s\gamma\gamma$ couplings ($|C_{s\gamma\gamma}/SM|\lesssim0.3$).
      Combining with Fig.\ref{fig3} we know these samples belong to the large-diphoton scenario.
  \item All samples with small $\eta$ ($\lesssim0.3$) have both small $sgg$ and $s\gamma\gamma$ couplings ($|C_{sgg}/SM|\lesssim0.5$ and $|C_{s\gamma\gamma}/SM|\lesssim0.6$).
      Combining with Fig.\ref{fig3} we know these samples belong to the small-diphoton scenario.
  \item All samples with large $|\sin\theta_S|$ ($\gtrsim0.4$) have both small $sgg$ couplings ($|C_{sgg}/SM|\lesssim0.5$) but large $s\gamma\gamma$ couplings ($|C_{s\gamma\gamma}/SM|\gtrsim0.5$).
      Combining with Fig.\ref{fig3} we know these samples also belong to the small-diphoton scenario.
\end{itemize}

\section{conclusion}
In this letter, motived by
the interesting topic of whether an additional scalar exists beyond the SM, and our unsureness of physics beyond the SM in the low mass region, especially the inconsistent results at around 95 GeV by Atlas and CMS collaborations in searching for light resonances in diphoton channel,
we study a light scalar in new physics models to interpret the different results in different parameter space, and further distinguishing them at the LHC.
We consider this idea in the Minimal Dilaton Model, which extends the SM by a dilaton/Higgs-like singlet scalar and a vector-like top partner.
In the calculations, we consider the theoretical constraints from vacuum stability and Landau pole, experimental constraints from EWPD, latest Higgs data at Run I and Run II of the LHC, and low-mass Higgs/resonances search at the LEP, Tevatron and LHC.
With the surviving samples under these constraints, we sort them into two scenarios:
the large-diphoton scenario (with $\sigma_{\gamma\gamma}/SM\gtrsim0.2$) and the small-diphoton scenario (with $\sigma_{\gamma\gamma}/SM\lesssim0.2$), which are favored by the CMS and Atlas results respectively.

We compare the two scenarios, check the characteristics in model parameters, scalar couplings, production and decay, and consider further distinguishing them at colliders.
Finally, we get the following conclusions:
\begin{itemize}
\item The large-diphoton scenario usually has small Higgs-dilaton mixing angle ($|\sin\theta_S|\lesssim0.2$) and small dilaton VEV $f$ ($0.5\lesssim\eta\equiv v/f\lesssim1$), while the small-diphoton scenario usually has large mixing ($|\sin\theta_S|\gtrsim0.4$) or large VEV ($\eta\equiv v/f\lesssim0.3$).
\item The former usually predicts a small $s\gamma\gamma$ coupling ($|C_{s\gamma\gamma}/SM|\lesssim0.3$) and a large $sgg$ coupling ($0.6\lesssim|C_{sgg}/SM|\lesssim1.2$), while the later usually predicts small $sgg$ coupling ($|C_{sgg}/SM|\lesssim0.5$).
\item The former can interpret the small diphoton excess by CMS at its central value, when $m_s\simeq95\GeV$, $\eta\simeq0.6$ and $|\sin\theta_S|\simeq0$.
\item The former usually predicts a negative correlation between Higgs couplings $|C_{h\gamma\gamma}/SM|$ and $|C_{hgg}/SM|$, while the later usually predicts the two couplings both smaller than 1, or $|C_{h\gamma\gamma}/SM|\lesssim0.9 \lesssim|C_{hgg}/SM|$.
\end{itemize}
The two scenarios can also be checked via $s\to b\bar{b}$ channel with $s$ produced through VBF or Vs at the LHC, and $s\to gg$ at future electron-positron colliders, where the loop effect of top quark sector in the scalar production may need to be considered, and we leave this study in our future work.

\section*{Acknowledgement}
We thank the helpful discussions with Dr. Tim Stefaniak on HiggsBounds and HiggsSignal.
This work was supported in part by the National Natural Science Foundation of China (NNSFC) under grant Nos. 11605123, 11547103, 11674253, 11547310.
JZ also thanks the support of the China Scholarship Council (CSC) under Grant No.201706275160 while he was working at the University of Chicago as a visiting scholar, and thanks the support of the US National Science Foundation (NSF) under Grant No. PHY-0855561 while he was working at Michigan State University.

\end{spacing}
\end{CJK*}
\end{document}